\documentclass[preprint,showpacs,amsmath,amssymb]{revtex4}

\usepackage{graphicx}
\usepackage{dcolumn}
\usepackage{bm}

\begin{document}

\title{Microwave Phase Detection at the Level of $10^{-11}$ rad}
\author{Eugene N. Ivanov}
\email{eugene@physics.uwa.edu.au}
\author{Michael E. Tobar} 
\email{mike@physics.uwa.edu.au}
\affiliation{University of Western Australia, School of
Physics M013, 35 Stirling Hwy., Crawley 6009 WA, Australia}

\date{\today}

\begin{abstract}
We report on a noise measurement system with the highest spectral resolution ever achieved in the microwave domain. It is capable of detecting the phase fluctuations of $rms$ amplitude of  $2\times10^{-11} rad/\sqrt{Hz}$ at Fourier frequencies above few $kHz$. Such precision allows the study of intrinsic fluctuations in various microwave components and materials, as well as precise tests of fundamental physics. Employing this system we discovered a previously unknown phenomenon of down-conversion of pump oscillator phase noise into the low-frequency voltage fluctuations. 
\end{abstract}

\maketitle

At frequencies, $\nu$, much less than a few THz at room temperature ($\nu<<k_BT_{0}/h$) , Nyquist thermal noise\cite{Nyquist, Oliver} restricts the resolution of phase (amplitude) measurements. For the conventional interferometric measurement system \cite{ITW98} containing a test sample, this limit is set by the Standard Thermal Noise Limit (STNL), given by $\sqrt{2 k_B T_{0}/(P_{inc}\alpha)}$ $rad/\sqrt{Hz}$ where, $k_B$ is the BoltzmannÕs constant, $T_{0}$ is the ambient temperature,  $\alpha$ is the loss in the test sample, $P_{inc}$ is the power incident on the test sample, and $h$ is PlankÕs constant. This work describes a new microwave noise measurement system with a phase noise floor of $2\times10^{-11} rad/\sqrt{Hz}$, which is $10 dB$ below the STNL and more sensitive than any laser interferometer operating at the same level of signal power\cite{laser02, laser98}.

\begin{figure}
\begin{center}
\includegraphics[width=3.0in]{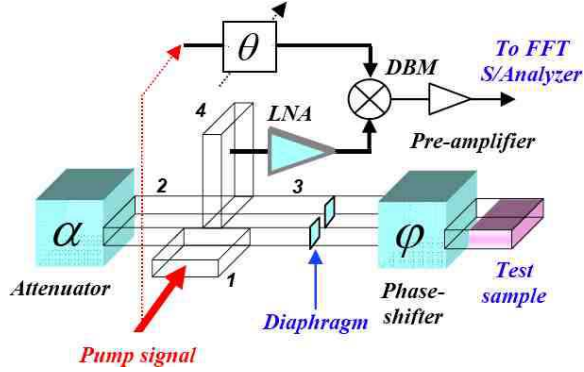}
\caption{Microwave noise measurement system based on an interferometer with a distributed resonator} \label{fig:nms}
\end{center}
\end{figure}
 
A schematic diagram of the noise measurement system is shown in Fig. \ref{fig:nms}. It consists of a standing wave interferometer based on a Magic Tee waveguide coupler and a microwave readout which includes a low-noise microwave amplifier ($LNA$), reference phase-shifter and a double-balanced mixer ($DBM$). The distinctive feature of the measurement system in Fig. \ref{fig:nms}  is a distributed resonator formed by an inductive diaphragm and a short-circuited piece of waveguide with a variable phase-shifter. A test sample can be either inserted directly into the waveguide or (if it is a coaxial device) coupled to it with an adapter. The smaller the loss in the test sample, the higher the sensitivity due to the extended lifetime of the distributed resonator. The signal reflected from the distributed resonator interferes destructively with a fraction of the incident signal at the interferometerÕs "dark port" (port 4 in Fig. \ref{fig:nms}). This cancels the carrier of the difference signal while preserving the noise modulation sidebands caused by non-thermal fluctuations in the test sample. The noise sidebands are amplified and converted into the voltage noise at the output of the $LNA$ and $DBM$ assembly. Depending on the setting of the reference phase shifter ($\theta$ in Fig. \ref{fig:nms})  the output voltage noise varies synchronously with either phase or amplitude fluctuations of the test sample.  The use of the LNA overcomes the relatively high technical fluctuations in the DBM making the effective noise temperature of the readout system $T_{RS}$ close to its physical temperature $T_{RS}\approx  T_{0}+T_{amp}$, where $T_{amp}$ is the effective noise temperature of the microwave $LNA$. 

The main reasons for choosing the waveguide components instead of micro-strip (used in \cite{ITW98, IT02}) were $(i)$ to increase the efficiency of power recycling (by reducing the distributed loss in the interferometer arms) and $(ii)$ to minimize the effect of technical noise sources inside the interferometer as some micro-strip components tend to exhibit an excess noise when exposed to relatively high levels of microwave power (of the order of $1 W$).

The phase sensitivity of the measurement system ($du/d\phi$ ) was optimized by adjusting the aperture of the inductive diaphragm and its distance from the symmetry plane of the Magic Tee. A piece of hollow waveguide approximately $10 cm$  long was initially used as a test sample. At $P_{inc} = 1W$  the highest value of phase sensitivity was measured to be $1.4 kV/rad$, corresponding to a factor of 4 enhancement relative to the conventional interferometer operating at the same level of input power.  This sensitivity enhancement was accompanied by more effective use of signal power, as only a few percent of the incident power was reflected from the input of the Magic Tee back to the pump source.

\begin{figure}
\begin{center}
\includegraphics[width=3.0in]{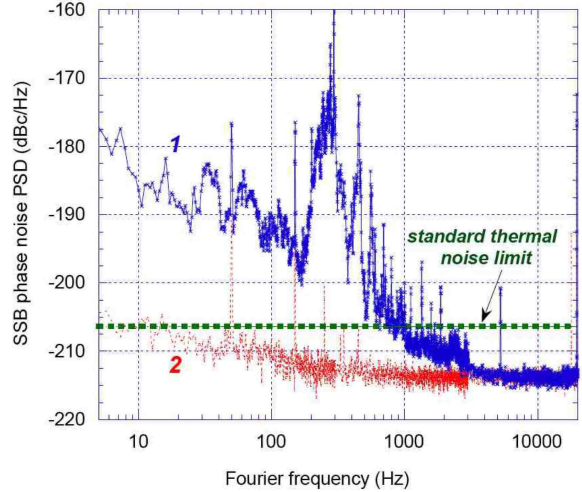}
\caption{SSB phase noise floors of a $9GHz$ measurement system; noise floor of the entire system (1); noise floor due to fluctuations in the microwave electronics of the readout system (2)} \label{fig:2}
\end{center}
\end{figure}

The Single Side Band (SSB) phase noise floor of the above measurement system is shown in Fig.\ref{fig:2} (curve 1). The rough peaks in the noise spectrum are of vibration and acoustic origin. At Fourier frequencies, $f>5 kHz$, the SSB phase noise floor approaches  $-213 dBc/Hz$. This is almost  $13 dB$ lower than the noise floor of a power recycled laser interferometer operating at the same level of signal power\cite{laser02, laser98}. Also, it should be noted that the above resolution was achieved in Òreal timeÓ, as opposed to ÒslowÓ measurements associated with the cross-correlation noise measurement technique first suggested in \cite{Walls} and later applied to interferometric measurements \cite{Ruby}. The horizontal line in Fig.\ref{fig:2} at the level $-203 dBc/Hz$  corresponds to the SSB STNL of the conventional interferometer calculated for a set of parameters typical for our experiments: $P_{inc} = 1W$, $T = 300K$ and $\alpha = 0.5 dB$ . As follows from Fig. \ref{fig:2}, the improvement $I$ in the resolution of noise measurement relative to the STNL is approximately $10 dB$. This is slightly less than what one would expect from the enhancement in the phase sensitivity $E$ mentioned earlier (factor of 4). This is because:  $I=E\sqrt{T_0/T_{RS}}$ , where, in our case, $T_{RS}\approx 390 K$.

The above measurements were repeated with the input of the $LNA$ terminated to evaluate the contribution of the microwave electronics to the overall uncertainty of the phase measurements. The resulting phase noise floor is given by curve 2 in Fig.\ref{fig:2}. The increasing divergence between two noise spectra at low Fourier frequencies could be attributed to intrinsic fluctuations inside the interferometer likely to be associated with the ambient temperature upsetting the interferometerÕs balance

\begin{figure}
\begin{center}
\includegraphics[width=3.0in]{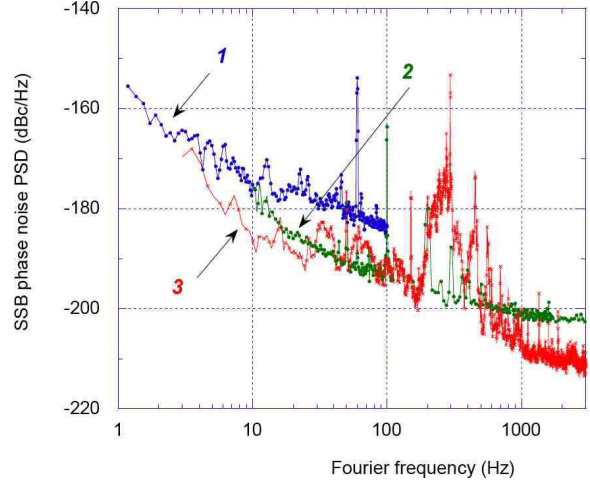}
\caption{Noise floors of different $9 GHz$ interferometric measurement systems: conventional coaxial interferometer (1); coaxial interferometer with power recycling (2); waveguide interferometer with signal recycling (3).} \label{fig:3}
\end{center}
\end{figure}

At low Fourier frequencies the spectral density of phase fluctuations seems to be independent on the type of components used to form the interferometer. Such conclusion could be derived from examining the phase noise floors of different interferometric measurement systems tested (Fig. \ref{fig:3}): $(i)$ conventional coaxial interferometer; $(ii)$ coaxial interferometer with recycling \cite{IT02} and $(iii)$ current waveguide based interferometer. At Fourier frequencies $f< 1 Hz$  all spectra in Fig. \ref{fig:3} converge to the same power law:  $\frac{1}{2} S_{\phi}\approx 3\times10^{-16}/f^2$ $rad^2/Hz$) which, we believe, is a ÒsignatureÓ of ambient temperature fluctuations. The above statement should be taken with some caution, as we did not take special measures to isolate our experimental setup from the vibration and acoustical influences.

A different situation was observed at $f > 10 Hz$ where the results of noise measurements with coaxial and micro-strip components were often irreproducible showing a large scatter from one experimental trial to another: in some cases the noise floor of the coaxial system was close to that of the waveguide system, while in others it was an order of magnitude higher. The noise performance of the waveguide based interferometers has always remained highly consistent. In the above experiments we used conventional continuously variable attenuators and phase-shifters, except for the waveguide interferometer, which was based on the precision rotary vane attenuator and phase shifter.

It should be noted that not only phase fluctuations in a test sample can be measured with the ultra-high precision: the measurements of amplitude fluctuations can equally benefit from the introduction of a distributed resonator. The point is that neither phase nor amplitude noise spectra are measured directly: they are both inferred from voltage noise at output of the measurement system when the latter is tuned either phase or amplitude sensitive. In case of the measurement system in Fig.\ref{fig:nms}, the spectral density of its output voltage noise was practically independent on the type of tuning. This meant equal contribution from both amplitude and phase fluctuations to the output voltage noise and, therefore, the possibility of their measurements with equal precision.

\begin{figure}
\begin{center}
\includegraphics[width=3.0in]{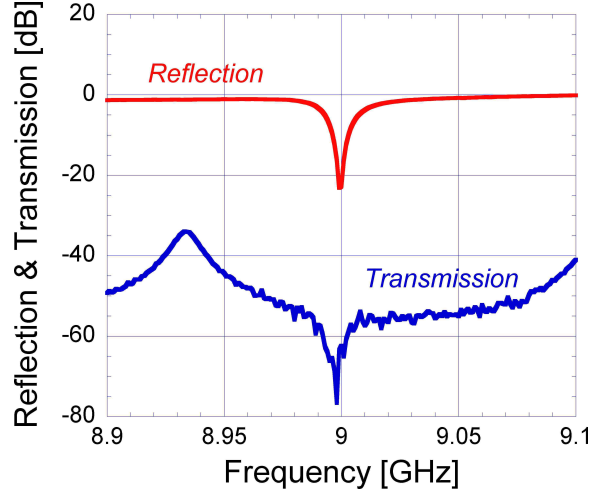}
\caption{Reflection and transmission coefficients of the microwave interferometer with power recycling} \label{fig:4}
\end{center}
\end{figure}

We also tested a microwave counterpart of the optical Michelson interferometer with power recycling. In such a case, an inductive diaphragm was removed from the side port of the Magic Tee and placed in the path of the incident signal. The aperture of the diaphragm and its distance from the symmetry plane of the Magic Tee were adjusted to minimise power of the reflected signal. When viewed with a microwave vector network analyser the input reflection coefficient of the interferometer resembled transfer function of a notch filter with the depth of 24 dB (Fig.\ref{fig:4}). At $P_{inc} = 1 W$ the maximum phase sensitivity of the power recycled interferometer was measured to be $0.9 kV/rad$.

The high resolution of the phase measurements observed in the above experiments was achieved by paying serious attention to technical noise sources. First, a careful balancing of the interferometer was required to avoid the excess $1/f$ voltage noise associated with the saturation of the $LNA$.  Secondly, the signal of the microwave pump source was band-pass filtered to remove the higher order harmonics from its spectrum. Such harmonics are not attenuated when the interferometer is balanced and can saturate the $LNA$.  Also, we have to deal with the enhanced sensitivity of the measurement system to pump oscillator frequency noise due to dispersion of the distributed resonator. The phase noise floor of the measurement system  due to the pump oscillator phase fluctuations  can be evaluated from: 

\begin{equation}
S_{\phi}^{n/f} = S_{\phi}^{osc}(f)(\frac{du/d\nu}{du/d\phi})^2 f^2.
\end{equation}

\begin{figure}
\begin{center}
\includegraphics[width=3.0in]{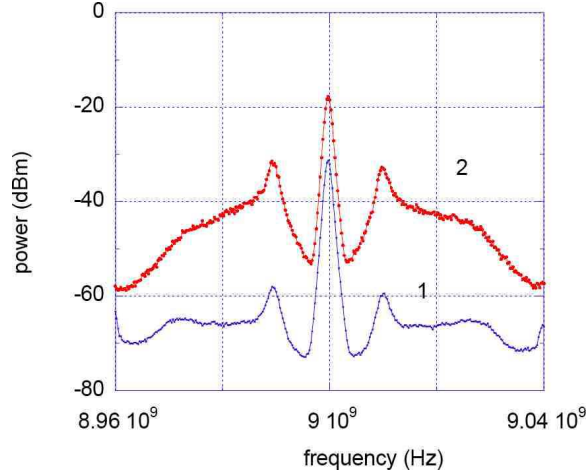}
\caption{$RF$ spectra at the output of the microwave $LNA$: $1$- conventional interferometer: $2$- interferometer with signal recycling
} \label{fig:5}
\end{center}
\end{figure}

Here $du/d\nu$ is the frequency sensitivity of the measurement system, which was measured to be $20 mV/kHz$ when the phase sensitivity was maximized ( i.e. $du/d\phi$  = $1.4 kV/rad$). Substituting the above values of sensitivities into (1) and assuming the phase noise of a high performance commercial frequency synthesizer, say E8257C from Agilent Technologies, one could conclude that the spectral resolution beyond the STNL would be only feasible at Fourier frequencies $f<3 kHz$. For this reason, most of our measurements (including the noise spectra in Fig. \ref{fig:2}) were performed with a composite source consisting of a low-phase noise $9 GHz$ oscillator based on Sapphire Loaded Cavity ($SLC$) resonator \cite{IT09} and a high power amplifier needed to generate a few Watts of useful power.  Within the range of Fourier frequencies $100 Hz$ to $200 kHz$ the phase noise of the composite source was dominated by the phase fluctuations of the high power amplifier. This, however, did not affect the resolution of noise measurements at $f < 700 kHz$. At higher Fourier frequencies the rise of oscillatorÕs phase noise (associated with the loss of gain of oscillator frequency control loop, as seen from Fig. \ref{fig:5}) was the main factor responsible for the loss of spectral resolution. It was also responsible for a tall peak around $10 MHz$ in the spectrum of mixer voltage noise. This high intensity voltage noise was saturating the pre-amplifier in front of the FFT spectrum analyzer (Fig. \ref{nms}) rising the ÒwhiteÓ noise floor of the entire measurement system \cite{IT09}.  Introduction of a passive low-pass filter in front of the pre-amplifier solved this problem.

We also discovered an additional type of noise inherent to narrow-band interferometers. It was a ÒwhiteÓ noise with spectral density weakly influenced by interferometerÕs amplitude mismatch, but showing strong dependence on interferometerÕs phase mismatch. Power at the interferometer "dark port" must be kept below approximately $50 dBm$ to prevent the loss of phase resolution due to this phenomenon. The origin of this noise was found to be related to the down-conversion of high frequency phase fluctuations of the pump oscillator signal. The point is that the narrow-band interferometer does not filter out fast phase fluctuations of the pump oscillator. This is illustrated by Fig. \ref{fig:5}, which shows the $RF$ spectra at the interferometer Òdark portÓ. The high intensity phase noise at Fourier frequencies of order $10 MHz$  is synchronously demodulated to $DC$ inside the $DBM$ when interacting with the replica of itself arriving at the second port of the DBM. We simulated these Ònoise-noiseÓ interactions by considering the propagation of a phase-modulated signal through the dispersive network. We also managed to induce this effect artificially by modulating frequency of the pump oscillator with a band limited random voltage noise centered around $10 MHz$. The problem of the excess noise in the recycled interferometer was solved by the narrow-band filtering the pump signal.  In our case, this was achieved by taking the output signal of the low-phase noise oscillator from the filtered port of the SLC resonator.

Concluding, we demonstrated the possibility of Òreal timeÓ noise measurements at microwave frequencies with the phase noise floor $10 dB$ below the STNL at Fourier frequencies above few kHz. This was achieved by combining the principles of microwave circuit interferometry with more efficient use of signal power and minimizing the influence of technical noise sources on the measurement process. We also discovered that under certain conditions an excess ÒwhiteÓ noise appears at the output of the narrow-band interferometer due to the self-demodulation of high frequency phase fluctuations in the spectrum of the incident signal. 

A narrow-band interferometer is ideally suited for studying the noise phenomena in low-loss components and materials at microwave frequencies. We used it to characterize the intrinsic phase fluctuations in ferrite circulators. So far, we could only claim that there is no general rule describing the phase noise in such devices: in some cases the circulator phase fluctuations were easily observable, while in others no noise was detected.

Recently, it was shown that a sensitive measurement of the isotropic Lorentz violation parameter, $\kappa_{tr}$, within the Standard Model Extension (SME)\cite{Kosto, Bailey}, could be attained by placing a magnetically permeable material inside the microwave interferometer \cite{Tobar}. Assuming a $10 cm$ long piece of a ferrite filled waveguide inserted in the recycled interferometer and the rotation rate of about $0.2 Hz$ \cite{Stanwix}, one can constrain $\kappa_{tr}$ within the interval of $\pm10^{-8}$ only after a single day of observations provided that no technical fluctuations are present in the ferrite \cite{Tobar}. This is about an order of magnitude better than current best limits \cite{Hoh, Reinhardt}. Because our measurements did not reveal any such fluctuations in magnetically shielded ferrite devices (including a ferrite loaded waveguide), the system looks promising for some future precise tests of the SME.

\begin{acknowledgments}
This work was funded by the Australian Research Council.
\end{acknowledgments}

\end{document}